\documentclass{article}%

\usepackage{amsmath}%
\usepackage{amsfonts}%
\usepackage{amssymb}%
\usepackage{graphicx}
\usepackage{hyperref}

\newcommand{\unit}[1]{\ensuremath{\, \mathrm{#1}}}
\newcommand{\sun}{\odot}

\begin{document}

\title{That is not dead which can eternal lie: the aestivation hypothesis for resolving Fermi's paradox}
\author{Anders Sandberg\footnote{Future of Humanity Institute, University of Oxford. Littlegate House, Suite 1, 16/17 St. Ebbe's Street. Oxford OX1 1PT. United Kingdom. \texttt{anders.sandberg@philosophy.ox.ac.uk}}, Stuart Armstrong\footnote{Future of Humanity Institute. \texttt{stuart.armstrong@philosophy.ox.ac.uk}}, Milan \'Cirkovi\'c\footnote{Future of Humanity Institute and Astronomical Observatory of Belgrade, Volgina 7, 11000 Belgrade, Serbia. \texttt{mcirkovic@aob.rs}}}
\maketitle

\begin{abstract}
\noindent If a civilization wants to maximize computation it appears rational to aestivate until the far future in order to exploit the low temperature environment: this can produce a $10^{30}$ multiplier of achievable computation. We hence suggest the ``aestivation hypothesis'': the reason we are not observing manifestations of alien civilizations is that they are currently (mostly) inactive, patiently waiting for future cosmic eras. This paper analyzes the assumptions going into the hypothesis and how physical law and observational evidence constrain the motivations of aliens compatible with the hypothesis. 
\end{abstract}

\noindent {\bf Keywords:} Fermi paradox, information physics, physical eschatology, extraterrestrial intelligence

\section{Introduction}
Answers to the Fermi question ("where are they?") typically fall into the groups ``we are alone'' (because intelligence is very rare, short-lived, etc.), ``aliens exist but not here'' (due to limitations of our technology, temporal synchronization etc.) or ``they're here'' (but not interacting in any obvious manner). This paper will examine a particular version of the third category, the aestivation hypothesis.

In previous work \cite{Spamming}, we showed that a civilization with a certain threshold technology could use the resources of a solar system to launch colonization devices to effectively every solar system in a sizeable part of the visible universe. While the process as a whole would last eons, the investment in each bridgehead system would be small and brief. Also, the earlier the colonization was launched the larger the colonized volume. These results imply that early civilizations have a far greater chance to colonize and pre-empt later civilizations if they wish to do so.

If these early civilizations are around, why are they not visible? The aestivation hypothesis states that they are aestivating\footnote{``Hibernation'' might be a more familiar term but aestivation is the correct one for sleeping through too warm summer months.} until a later cosmological era. 

The argument is that the thermodynamics of computation make the cost of a certain amount of computation proportional to the temperature. 

Our astrophysical and cosmological knowledge indicates that the universe is cooling down with cosmic time. Not only star formation within galaxies winds down and dies out on timescales of $10^9$ - $10^{10}$ yrs but even the cosmic background radiation temperature is becoming exponentially colder.

As the universe cools down, one Joule of energy is worth proportionally more. This can be a substantial ($10^{30}$) gain. Hence a civilization desiring to maximize the amount of computation will want to use its energy endowment as late as possible: using it now means far less total computation can be done. Hence an early civilization, after expanding to gain access to enough raw materials, will settle down and wait until it becomes rational to use the resources. We are not observing any aliens since the initial expansion phase is brief and intermittent and the aestivating civilization and its infrastructure is also largely passive and compact.

The aestivation hypothesis hinges on a number of assumptions that will be examined in this paper. The main goal is to find what physical and motivational constraints make it rational to aestivate and see if these can be met. If they cannot, then the aestivation hypothesis is not a likely answer for the Fermi question.

This paper is very much based on the physical eschatology research direction started by Freeman Dyson in \cite{Dyson} combined with the SETI approach dubbed ``Dysonian SETI'' \cite{DysonianSETI}. Physical eschatology attempts to map the long-term future of the physical universe given current knowledge of physics \cite{PhysEschatology,Dying}. This includes the constraints for life and information processing as the universe ages. Dysonian SETI takes the approach to widen the search to look for detectable signatures of highly advanced civilizations, in particular megascale engineering. Such signatures would not be deliberate messages, could potentially outlast their creators, and are less dependent on biological assumptions about the originating species (rather, they would stand out because of unusual physics compared to normal phenomena). 
%

It should be noted that in the physical eschatology context aestivation/hibernation has been an important issue for other reasons: Dyson suggested it as a viable strategy to stretch available energy and heat dissipation resources \cite{Dyson}, and Krauss and Starkman critiqued this approach \cite{Nothing}.


\section{Aestivation}

\begin{quote}
The Old Ones were, the Old Ones are, and the Old Ones shall be. Not in the spaces we know, but between them. They walk serene and primal, undimensioned and to us unseen.\\
H.P. Lovecraft, The Dunwich Horror and Others
\end{quote}

An explanation for the Fermi question needs to account for a lack of observable aliens given the existence of human observers in the present (not taking humans into account can lead to anthropic biases: we are not observing a randomly selected sample from all possible universes but rather a sample from universes compatible with our existence \cite{bostrom2013anthropic}). 

The aestivation hypothesis makes the following assmptions:

\begin{enumerate}
\item There are civilizations that mature much earlier than humanity. 
\item These civilizations can expand over sizeable volumes, gaining power over their contents. 
\item These civilizations have solved their coordination problems.
\item A civilization can retain control over its volume against other civilizations. 
\item The fraction of mature civilizations that aestivate is non-zero 
\item Aestivation is largely invisible.
\end{enumerate}

Assumption 1 is plausible, given planetary formation estimates \cite{Lineweaver}. Even a billion years of lead time is plenty to dominate the local supercluster, let alone the galaxy. If this assumption is not true (for example due to global synchronization of the emergence of intelligence \cite{Phase}) we have an alternative answer to the Fermi question.

We will assume that the relevant civilizations are mature enough to have mastered the basic laws of nature and their technological implications to whatever limit this implies. More than a million years of technological development ought to be enough.

Assumption 2 is supported by our paper \cite{Spamming}. In particular, if a civilization can construct self-replicating technological systems there is essentially no limit to the amount of condensed matter in solar systems that can be manipulated. These technological systems are still potentially active in the current era. Non-expanding aestivators are entirely possible but would not provide an answer to the Fermi question. 

If assumption 2 is wrong, then the Fermi question is answered by a low technology ceiling and that intelligence tends to be isolated from each other by distance. 

Assumption 3: This paper will largely treat entire civilizations as the actors rather than their constituent members.
In this regard they are ``singletons'' in the sense of Nick Bostrom: having a highest level of agency that can solve internal coordination problems \cite{Singleton}. Singletons may be plausible because they provide a solution to self-induced technological risks in young technological civilizations (preventing subgroups from accidentally or deliberately wiping out the civilization). Anecdotally, humanity has shown an increasing ability to coordinate on ever larger scales thanks to technological improvements (pace the imperfections of current coordination). In colonization scenarios such as \cite{Spamming} coordinating decisions made before launch could stick (especially if they have self-enforcing aspects) even when colonies are no longer in contact with each other. However, the paper does not assume all advanced civilizations will have perfect coordination, merely very good coordination. 

Assumption 4 needs to handle both civilizations emerging inside the larger civilization and encounters with other mature civilizations at edges of colonization. For the first case the assumption to some degree follows from assumption 2: if desired the civilization could prevent misbehavior of interior primitive civilizations (the most trivial way is to prevent their emergence, but this makes the hypothesis incompatible with our existence \cite{Berzerker}). Variants of the zoo hypothesis work but the mature civilization must put some upper limit to technological development or introduce the young civilizations into their system eventually. The second case deals with civilizations on the same level of high technological maturity: both can be assumed to have roughly equal abilities and resources. If it can be shown that colonized volumes cannot be reliably protected against invasion, then the hypothesis falls. Conversely, if it could be proven that invasion over interstellar distances is logistically unfeasible or unprofitable, this could be regarded as a (weak) probabilistic support for our hypothesis.

Assumption 5 is a soft requirement. In principle it might be enough to argue that aestivation is possible and that we just happen to be inside an aestivation region. However, arguing that we are not in a typical galaxy weakens the overall argument. More importantly, the assumption explains the absence of non-aestivating civilizations. Below we will show that there are strong reasons for civilizations to aestivate: we are not arguing that all civilizations must do it but that it is a likely strategy. Below we will discuss the issues of civilizational strategies and discounting that strengthen this assumption. 

Assumption 6 is the key assumption for making the hypothesis viable as a Fermi question answer: aestivation has to leave the universe largely unchanged in order to be compatible with evidence. The assumption does not imply deliberate hiding (except, perhaps, of the core part of the civilization) but rather that advanced civilizations do not dissipate much energy in the present era, making them effectively stealthy. The main empirical prediction of the hypothesis is that we should observe a strong dampening of processes that reduce the utility for far future computation.

\section{Civilizational goals} 


While it may be futile to speculate on the values of advanced civilizations, we would argue that there likely exists convergent instrumental goals. 

What has value? Value theories assign value to either states of the world or actions. 

State value models require resources to produce high-value states. If happiness is the goal, using the resources to produce the maximum number of maximally happy minds (with a tradeoff between number and state depending on how utilities aggregate) would maximize value. If the goal is knowledge, the resources would be spent on processing generating knowledge and storage, and so on. For these cases the total amount of produced value increases monotonically with the amount of resources, possibly superlinearly.




Actions also require resources and for actions with discernible effects the start and end states must be distinguishable: there is an information change. Even if value resides in the doing, it can be viewed as information processing. 

The assumption used in this paper is that nearly all goals benefit from more resources and that information processing and storage are good metrics for measuring the potential of value creation/storage. They are not necessarily final goals but instrumentally worth getting for nearly all goals for nearly all rational agents \cite{SuperWill,Bradbury06}. 




As we will see below, maintaining communication between different parts of an intergalactic civilization becomes impossible over time. If utility requires contact (e.g. true value is achieved by knowledge shared across a civilization) then these limits will strongly reduce the utility of expanding far outside a galactic supercluster. If utility does not require contact (e.g. the happiness or beauty of local entities is what is valuable in itself) then expanding further is still desirable.

There may also be normative uncertainty: even if a civilization has a convincing reason to have certain ultimate values, it might still regard this conclusion as probabilistic and seek to hedge its bets. Especially if the overall plan for colonization needs to be formulated at an early stage in its history and it cannot be re-negotiated once underway it may be rational to ensure that alternative value theories (especially if easily accommodated) can be implemented. This may include encountered alien civilizations, that might possibly have different but valid ultimate goals.


\section{Discounting}

Another relevant property of a civilization is the rate of its temporal discounting. How much is the far future worth relative to the present? There are several reasons to suspect advanced civilizations have very long time horizons.

In a dangerous or uncertain environment it is rational to rapidly discount the value of a future good since survival to that point is not guaranteed. However, mature expanding civilizations have likely reduced their existential risks to a minimum level and would have little reason to discount strongly (individual members, if short-lived, may of course have high discount rates). More generally, the uncertainty of the future will be lower and this also implies lower discount rates. 

Specific policies that take a long time to implement such as megascale engineering or interstellar travel also favor low discount rates. There could be a selection effect here: high discount rates may prevent such long-range projects and hence only low-discount rate civilizations will be actors on the largest scales. 

It also appears likely that a sufficiently advanced civilization could regulate its ``mental speed'', either by existing as software running on hardware with a variable clock speed, or by simply hibernating in a stable state for a period. If this is true, then the value of something after a period of pause/hibernation would be determined not by the chronological external time but how much time the civilization would subjectively experience in waiting for it. Changes in mental speed can hence make temporally remote goods more valuable if the observer can pause until they become available and there is no alternative cost for other goods.

This is linked to a reduction of opportunity costs: advanced civilizations have mainly ``seen it all'' in the present universe and do not gain much more information utility from hanging around in the early era\footnote{Some exploration, automated or crewed, might of course still occur during the aestivation period, to be reported to the main part of the civilization at its end.}

There are also arguments that future goods should not be discounted in cases like this. What really counts is fundamental goods (such as well-being or value) rather than commodities; while discounting prices of commodities makes economic sense it may not make sense to discount value itself \cite{Broome}.  

This is why even a civilization with some temporal discounting can find it rational to pause in order to gain a huge reward in the far future. If the subjective experience is an instant astronomical multiplication of goods (with little risk) it is rational to make the jump. However, it requires a certain degree of internal coordination to overcome variation in individual time preferences (hence assumption 3).

\section{Thermodynamics of extreme future computation}


Computation is a fundamentally physical process, tied into thermodynamic considerations. 

Foremost for our current purposes is Landauer's limit:
at least 
\begin{equation}
E \geq kT \ln(2) \unit{J}
\end{equation}
 need to be dissipated for an irreversible change of one bit of information \cite{Landauer}. Logically irreversible manipulation of information must be accompanied by an entropy increase somewhere.

It should be noted that the thermodynamic cost can be paid by using other things than energy. An ordered reservoir of spin \cite{Spin} or indeed any other conserved quantity \cite{Conserved} can function as payment. The cost is $\ln(2)/\lambda$ where $\lambda$ is related to the average value of the conserved quantity. However, making such a negentropy reservoir presumably requires physical work unless it can be found naturally untapped.

\subsection{Reversible and quantum computation}

There exists an important workaround for the Landauer limit: logically reversible computations do not increase entropy and can hence be done in principle without any need of energy dissipation. 

It has been shown that any logically irreversible computation can be expressed as a reversible computation by storing all intermediate results, outputting the result, and then retracing the steps leading up to the final state in reverse order, leaving the computer in the original state. The only thermodynamic costs would be setting the input registers and writing the output \cite{Bennett}. More effective reversible implementation methods are also known, although the number of elementary gates needed to implement a $n$-bit irreversible function scales somewhere between $n2^n/\log n$ and $n2^n$ \cite{Synthesis}. 


Quantum computation is also logically reversible since quantum circuits are based on unitary mappings. While the following analysis will be speaking of bits, the actual computations might occur among qubits and the total reversible computational power would be significantly higher. 

If advanced civilizations do all their computations as reversible computations, then it would seem unnecessary to gather energy resources (material resources may still be needed to process and store the information). However, irreversible operations must occur when new memory is created and in order to do error correction. 

In order to create $N$ zeroed bits of memory at least $kT\ln(2)N \unit{J}$ have to be expended, beside the work embodied in making the bit itself. 

Error correction can be done with arbitrary fidelity thanks to error correcting codes but the actual correction is an irreversible operation. The cost can be deferred by transferring the tainted bit to an ancilla bit but in order to re-use ancillas they have to be reset. 

Error rates are suppressed by lower temperature and larger/heavier storage. Errors in bit storage occur due to classical thermal noise (with a probability proportional to $e^{-E_b/kT}$ where $E_b$ is the barrier height) and quantum tunneling (probability approximately $e^{-2r\sqrt{2mE_b}/\hbar}$ where $m$ is the mass of the movable entity storing the bit and $r$ is the size of the bit). The minimum potential height compatible with any computation is 
\begin{equation}
E_b^{min} \approx kT\ln(2) + \hbar (\ln(2))^2/8mr^2
\end{equation}
\cite{Gedanken}. 

If a system of size $R$ has mass $M$ divided into $N$ bits, each bit will have size $\approx R/N^{1/3}$. If a fraction of the mass is used to build potential barriers, we can approximate the height of the barrier $E_b$ in each bit as the energy in all the covalent bonds in the barrier. If we assume diamond as a building material, the total bond energy is $1.9\unit{kJ/mol} = 1.6 \cdot 10^{5} \unit{J/kg}$.

Using a supercluster $r=50 \unit{Mpc}$, $M=10^{43} \unit{kg}$, it can be subdivided into $10^{61}$ bits without reaching the limit given the current 3K background temperature. For the future $T_{\mathrm{dS}}$ temperature (see below) the limit is instead near $10^{75}$ bits (each up to 13 cm across); here the limiting factor is tunneling (the shift occurs around $T=10^{-8} \unit{K}$). However, this would require bits far lighter than individual atoms or even electrons\footnote{In theory neutrinos or very light WIMPs such as axions could fulfill the role but there is no known mechanism for confining them in such a way that they could store information reliably.}. In practice, the ultimate limiting capacity due to matter chunkiness would be on the order of $10^{69}$ bits. However, these ``heavy'' bits would have 15 orders of magnitude of safety margin relative to the potential height and hence have minuscule tunneling probabilities. So while error correction and repair will have to be done (as noted by Dyson, over sufficiently long timescales ($10^{65}$ years) matter is a liquid), the rate can be very low. 

Quantum computing is also affected by environmental temperature and quantum error correction can only partially compensate for this \cite{Cafaro}.
In theory there might exist topological quantum codes that are stable against finite temperature disturbances 
\cite{Hamma} but again there might be thermal or tunneling events undermining the computing system hardware itself.

Hence, even civilizations at the boundary of physical feasibility will have to perform some dissipative computational operations. They cannot wait indefinitely (since slow cosmological processes will erode their infrastructure and data) and will hence eventually run out of energy. 

\subsection{Cooling}
While it is possible for a civilization to cool down parts of itself to any low temperature, the act of cooling is itself dissipative since it requires doing work against a hot environment. The most efficient cooling merely consists of linking the computation to the coldest heat-bath naturally available. In the future this will be the cosmological background radiation\footnote{The alternative might be supermassive black hole horizons at $T=\hbar c^3/8 \pi G M k$, which in the present era can be far colder than the background radiation \cite{opatrny2017life}. Supermassive $10^9 M_\sun$ black holes will become warmer than the background radiation in 520 Gyr (assuming constant mass).}, which is also conveniently of maximal spatial extent. 

The mean temperature of the background radiation is redshifted and declines as $T(t)=T_0/a(t)$ where $a(t)$ is the cosmological scale factor \cite{CMB}. Using the long term de Sitter behavior $a(t) = e^{Ht}$ produces $T(t) = T_0 e^{-Ht}$. This means that one unit of energy will be worth an exponentially growing amount of computations if one waits long enough.

However, the background radiation is eventually redshifted below the constant de Sitter horizon radiation temperature $T_{\mathrm{dS}} = \sqrt{\Lambda/12\pi^2} = H/2\pi \approx 2.67\cdot 10^{-30}$ K \cite{CMB,Nothing}. 
This occurs at time $t=H \ln(T_0/T_{\mathrm{dS}})$, in about $1.4\cdot 10^{12}$ years. 
There will be some other radiation fields (at this point there are still active stars) but the main conclusion is that there is a final temperature of the universal heat bath. This also resolves a problem pointed out in \cite{Nothing}: in open universes the total number of photons eventually received from the background radiation is finite and all systems decouple thermally from it. In this case this never happens.

One consequence is that there will always be a finite cost to irreversible computations: without infinite resources only a finite number of such computations can be done in the future of any civilization.

This factor also avoids the ``paradox of the indefinitely postponed splurge'', where if it is always beneficial to postpone exploitation. Hence, it is rational at some point for aestivating civilizations to start consuming resources.
The limiting temperature of $T=10^{-8} \unit{K}$ where error correction stops being temperature-limited and instead becomes quantization-limited might be another point where it is rational to start exploitation (this will occur in around 270 billion years). Had the temperature decline been slower, the limit might have been set by galactic evaporation ($10^{16}$ years) or proton decay (at least $10^{34}$ years).

\subsection{Value of waiting}

A comparison of current computational resources to late era computational resources hence suggest a potential multiplier of $10^{30}$!

Even if only the resources available in a galactic supercluster are exploited, later-era exploitation produces a payoff far greater than any attempt to colonize the rest of the accessible universe and use the resources early. In fact, the mass-energy of just the Earth itself ($5.9\cdot 10^{24} \unit{kg}$) would be more than enough to power more computations than could currently be done by burning the present observable universe! ($6\cdot 10^{52} \unit{kg}$)\footnote{This might suggest that ``stay at home'' civilizations might hence abound, content to wait out the future in their local environment since their bounded utility functions can be satisfied eventually. Such civilizations might be common but they are not a solid explanation for the Fermi question since their existence does not preclude impatient (and hence detectable) civilizations. However, see \cite{Parkinson}. In addition, this strategy is risky since expansive civilizations may be around.}

In practice the efficiency will be lower but the multiplier tends to remain astronomical.

A spherical blackbody civilization of radius $r$ using energy at a rate $P$ surrounded by a $T_{\mathrm{dS}}$ background will have an equilibrium temperature (neglecting external and internal sources) 
\begin{equation}
T = [P/(4 \pi \sigma r^2) + T_{\mathrm{dS}}^4]^{1/4}.
\end{equation}
For a $r=50 \unit{Mpc}$ super-cluster sized civilization this means that maintaining a total power of $P=1 \unit{W}$ would keep it near $2.8\cdot 10^{-11} \unit{K}$. At this temperature the civilization could do $3.8\cdot 10^{33}$ irreversible computations per second. The number $P/kT\ln(2)$ bit erasures per second increases as $P^{3/4}$. At first this might suggest that it is better to ignore heat and use all resources quickly. However, a civilization with finite energy $E$ will run out of it after time $E/P$ and the total amount of erasures will then be $E/kT\ln(2)$: this declines as $P^{-1/4}$. Slow energy use produces a vastly larger computational output if one is patient.

As noted by Gershenfeld, optimal computation needs to make sure all internal states are close to the most probable state of the system, since otherwise there will be extra dissipation \cite{Gershenfeld}. Hence there is a good reason to perform operations slowly. Fortunately, time is an abundant resource in the far future. In addition, a civilization whose subjective time is proportional to the computation rate will not internally experience the slowdown. 

The Margolus-Levitin limit shows that it takes at least time $\pi \hbar / 2 E$ to move to a new orthogonal state \cite{Margolus}. For $E=kT_{\mathrm{dS}}$ this creates a natural ``clock speed'' of $3.8\cdot 10^{11}$ years. However, some of the energy will be embodied in computational hardware; even if the hardware is just single electrons the clock speed would be $2.0\cdot 10^{-21} \unit{s}$: this particular limit is not a major problem for this style of future computation. 

A stronger constraint is the information transmission lags across the civilization. For this example the time to send a message across the civilization is 326 million years. Whether this requires a very slow clock time or not depends on the parallelizability of the computations done by the civilization, which in turn depends on its values and internal structure. Civilizations with more parallelizable goals such as local hedonic generation would be able to have more clock cycles per external second than more ``serial'' civilizations where a global state needs to be fully updated before the next step can be taken. However, external time is of little importance given the paucity of external events. Even if proton decay in $10^{34}$ years puts a final deadline to the computation, this would still correspond to $10^{25}$ clock cycles.

Heat emission at very low temperature is also a cause of slowdown. The time to radiate away the entropy of a single bit erasure scales as $t_{\mathrm{rad}}=k\ln(2)/4\pi\sigma r^2 T^3$. For a 50 Mpc radius system this is $5.6\cdot 10^{-66} T^{-3} \unit{s}$: for $10^{-8}$ K the time is on the order of $10^{-42}$ s but at $10^{-30}$ K it takes $10^{17}$ years.

If the civilization does have a time limit $t_{\mathrm{max}}$, then it is rational to use $P=E/t_{\mathrm{max}}$ and the total number of operations will be proportional to $t_{\mathrm{max}}^{1/4}$. Time-limited civilizations do have a reason to burn the candle at both ends. Time-limited civilizations still gain overall computational success by waiting until the universe cools down enough so their long-term working temperature $T$ is efficient. At least for long time limits like proton decay a trillion years is a short wait.

The reader might wonder whether starting these computations now is rational since the universe is quickly cooling and will soon (compared to the overall lifespan of the civilization) reach convenient temperatures. The computational gain of doing computations at time $t$ is $\propto \exp(Ht)$: it increases exponentially until the temperature is dominated by the internal heating rather than the outside temperature. Since most of the integrated value accrues within the last e-folding and the energy used early was used exponentially inefficiently, it is not worth starting early even if the wait is a minuscule fraction of the overall lifespan.

A civilization burning through the baryonic mass of a supercluster before proton decay in $10^{33}$ years has a power of $5.7\cdot 10^{21}$ W (similar to a dim red dwarf star) and a temperature of $7.6\cdot 10^{-6}$ K, achieving $10^{80}$ erasures. The most mass-limited version instead runs close to $10^{-30}$ K, has a power of somewhere around $10^{-75}$ W and achievs $10^{115}$ erasures -- but each bit erasure, when it happens causes a 100 quadrillion year hiatus. A more realistic system (given quantization constraints) runs at $10^{-8}$ K and would hence have power $10^{10}$ W (similar to a large present-day power plant), running for $5.7\cdot 10^{44}$ years and achieving $10^{93}$ erasures.

\section{Resources}

The amount of resources available to advanced civilizations depends on what can ultimately be used. Conservatively, condensed molecular matter such as planets and asteroids are known to be useful for both energy, computation, and information storage. Stars represent another form of high density matter that could plausibly be exploited, both as energy sources and material. Degenerate stars (white and black dwarfs, neutron stars) are also potential high density resources.
Less, conservatively, 
there are black holes, from which mass-energy can be extracted \cite{Unruh,Lawrence} (and, under some conditions, can act as heat sinks as mentioned above). 
Beyond this, there is interstellar gas, intergalactic gas, and dark matter halos.

For the purposes of this paper we will separate the resources into energy resources that can power computations and matter resources that can be used to store information, process it or (in a pinch) be converted into energy. Mass-energy diffused as starlight or neutrinos, and stars lost from galaxies are assumed to have become too dilute to be useful. Dark energy appears to be unusable in principle. Dark matter may be useful as an energy resource by annihilation even if it cannot sustain information processing structures.

Not all resources in the universe can be reached and exploited. The lightspeed limit forces civilizations to remain within a light-cone and the accelerating expansion further limits how far probes can be sent. Based on the assumptions in \cite{Spamming} the amount of resources that can be reached within a 100 Mpc supercluster or by traveling at 50\%, 80\% or 99\% c are listed in table \ref{tab:Resources}.
	
\begin{table}
	\centering
	\small
		\begin{tabular}{p{3.5cm}llll}
										& 100 Mpc		&	1.24 Gpc &	2.33 Gpc & 4.09 Gpc \\
\hline
Planetary bodies
and condensed matter.& $3.92 \cdot 10^{42}$
& $7.47 \cdot 10^{45}$
& $4.96 \cdot 10^{46}$
& $2.68 \cdot 10^{47}$
\\
Stars (including 
white dwarfs, 
neutron stars and 
substellar objects)	& $2.74 \cdot 10^{45}$
& $5.23 \cdot 10^{48}$
& $3.47 \cdot 10^{49}$
& $1.88 \cdot 10^{50}$
\\
Black holes					& $8.25 \cdot 10^{43}$
& $1.57 \cdot 10^{47}$
& $1.04 \cdot 10^{48}$
& $5.65 \cdot 10^{48}$
\\
Interstellar gas		& $8.70 \cdot 10^{44}$
& $1.66 \cdot 10^{48}$
& $1.10 \cdot 10^{49}$
& $5.95 \cdot 10^{49}$
\\
Intergalactic gas		& $4.66 \cdot 10^{46}$
& $8.89 \cdot 10^{49}$
& $5.90 \cdot 10^{50}$
& $3.19 \cdot 10^{51}$
\\
Dark matter					& $2.81 \cdot 10^{47}$
& $5.36 \cdot 10^{50}$
& $3.55 \cdot 10^{51}$
& $1.92 \cdot 10^{52}$
\\
Total								& $3.31 \cdot 10^{47}$
& $6.31 \cdot 10^{50}$
& $4.19 \cdot 10^{51}$
& $2.27 \cdot 10^{52}$
\\
		\end{tabular}
	\caption{Resources available to civilizations expanding at different speeds, or within a single supercluster. Estimates based on \cite{Planck} and \cite{Inventory}. Distances measured in co-moving coordinates, mass in kilograms.
	}
	\label{tab:Resources}
\end{table}









\subsection{Changes over time}

\subsubsection{Stellar fusion}

Stellar lifetime energy emissions are proportional to mass (with high luminosity stars releasing more faster), $E_{\mathrm{life}} = LT = L_\sun (M/M_\sun) 3.16 \cdot 10^{17} \unit{J}$, leading to a lifetime mass loss through energy emission of 
\begin{equation}
M_{\mathrm{loss}}= 1.35\cdot 10^{27} (M/M_\sun) \unit{kg} = 6.79 \cdot 10^{-4} M_\sun (M/M_\sun).
\end{equation}
Lighter stars loose less mass but since the mean mass star is $0.7 M_\sun$ this is on the order of a typical value. 

Hence stellar fusion is not a major energy waste if mass can be converted into energy. This is important  for the aestivation hypothesis: had stellar fusion been a major source of mass loss it would had been rational to stop it during the aestivation period, or at least to gather up the lost energy using Dyson shells: both very visible activities that can be observationally ruled out (at least as large-scale activities).

Fusion processes also produce nuclei that might be more useful for computation without any need for intervention. Long term elemental mass fractions approach 20\% hydrogen, 60\% helium and 20\% other elements over $10^{12}$ year timescales \cite{Dying}.

\subsubsection{Black hole formation}
Stellar black holes permanently reduces the amount of baryonic matter available even if their mass-energy is exploitable. At present a mass fraction $\approx 2.5\%$ of the star formation budget is lost this way \cite{Inventory}.

To prevent this star formation of masses above $25M_\sun$ needs to be blocked. This could occur by interventions that cause extra fragmentation of clouds condensing into heavy stars. Adding extra dust to the protostellar cloud could induce rapid cooling and hence fragmentation. 
This would require dust densities on the order of $10^{-5}\rho_{\mathrm{gas}}$ and for a stellar formation rate $\approx 1M_\sun$ per year would require seeding galactic clouds with $\approx 10^{-5}M_\sun$ per year. 
Average nucleosynthetic yields are $\approx 0.0025$, so there would be enough metals produced per year to seed the clouds by about two orders of magnitude.

Other ways of inducing premature fragmentation might be to produce radiation bursts (from antimatter charges or directed illumination from Dyson-shell surrounded stars) or pressure waves, while magnetic fields might slow core collapse.
The energies involved would be of the order of the cloud potential energy; for a Bok globule this might require $(3G/5)(50M_\sun^2/1 \unit{ly})=4.2\cdot 10^{36} \unit{J}$, about $10^3$ sun-years of luminosity. 
Given that cloud collapse is a turbulent process it might be possible to prevent massive star formation through relatively energy-efficient chaos control.

While these methods might be invisible over long distances their effects ought to be noticeable due to a suppression of starbursts and heavy blue-white stars.

\subsubsection{Galactic winds}

While stars can lose significant amount of mass by outflows, this gas is recycled through the interstellar medium into new stars.
However, some of this medium may be lost due to galactic winds and may hence become long-term inaccessible. Galactic winds are likely proportional to the stellar formation rate  ($M'/SFR$ around 0.01-10) plus contributions from active galactic nuclei, and may have significantly depleted early galaxies. Starbursts can lose $10^5-10^6 M_\sun$ in dwarf galaxies and $10^8-10^{10}$ in ULIRGs, partially by entraining neutral matter. \cite{GalWind}
However, for large galaxies the actual escape fractions may be low due to dark matter halos keeping the gas bound and dampening the outflow through drag, keeping it below 4\% \cite{DynLim}. There can also be ongoing infall due to the halo that more than compensates for the loss. 



Due to the uncertainty about the wind budget it is not clear whether a civilization prioritizing baryonic matter might want to prevent galactic winds. A few approaches may be possible. One way, as described above, is to prevent too vigorous star formation. Such a civilization would also be interested in keeping the galactic nucleus quiescent, perhaps using the ``stellar billiards'' method suggested in the next section to manipulate orbits near the central black hole. These methods would likely be hard to detect, except for the reduction in hot stars\footnote{Another, somewhat extreme, approach would be to bag the galaxy: at a surface density of $0.77\cdot 10^{-6} \unit{kg/m^2}$ a graphene balloon encompassing a Milky Way-sized galaxy ($r=20$ kpc) would weigh $1.8\cdot 10^6 M_\sun$. Given a current carbon mass fraction of 0.0046 this would require mining $0.4\cdot 10^9 M_\sun$ stars or gas clouds, or fusing several million solar masses. Since the saved gas mass would be just one order of magnitude larger than the balloon weight, it is likely that this approach is not effective.}.


\subsubsection{Galaxy collisions}

Galactic ``splashes'' and mergers may lose significant amount of gas, stars and halos to intergalactic space. 



Can collisions be prevented? Typical galactic peculiar velocities are on the order of 200-300 km/s: in order to avoid a collision the velocity of one or both galaxies need to be changed by this amount by expelling some mass-energy. The rocket equation hence implies a final mass $m_1 = m_0 e^{-\Delta v/v_e}$. If $v_e \approx 1000 \unit{km/s}$, such as in the case of deliberate ejection of hypervelocity stars\footnote{This can be achieved in a step-wise process. First, Shkladov thrusters \cite{StellarEng} are built around many stars. While these thrusters are unable to change stellar velocities significantly over a Gyr, they are enough to engineer close encounters with passing stars within a few million years. This allows a bootstrap process where gravity assists can bring velocity changes of stars up to the average stellar velocities of 100 km/s as well as allow course corrections. Stars are then sent on slingshot orbits past the central black hole, ejecting them and accelerating the galaxy. This scheme mainly converts stellar kinetic and potential energy into thrust.}, the mass loss will be 10\%. 
For an optimal exhaust velocity of $c$ (using some form of mass-energy conversion) the loss is around 3.5\%. While this is a small fraction of the entire galactic mass, most of it ($>95\%$) is in the form of a dark matter halo: if only baryonic matter can be used to power the evasive maneuver practically all of it would be lost in the process. 

Hence, if a civilization seeks to retain significant amounts of baryonic matter it is not rational in general to prevent galactic collisions. Civilizations that instead care about overall mass-energy may be more motivated to prevent splashes of halo material by sacrificing some mass-energy in a very visible maneuver. If the maneuver takes one Gyr and uses mass-energy conversion, the average luminosity will be $3.5\% \cdot 10^{12}M_\sun c^2 /( 10^9 \unit{yr}) = 2\cdot 10^{41}\unit{W} = 5\cdot 10^{14} L_\sun$. Streams of $10^{10}$ hypervelocity stars would also likely be very noticeable.

\subsubsection{Expansion}

Given the current $\Lambda$CDM model of the universe, the expansion rate is increasing and approaching a de Sitter expansion. This leads to the eventual separation of all gravitationally bound systems from each other by insurmountable distances, leaving each an ``island universe'' within their cosmological horizon. The local group of galaxies will likely be separated from the Virgo supercluster and in 100 Gyr entirely separate \cite{NearbyLargeScale}.
Busha et al. find a criterion for structures remaining bound,
\begin{equation}
M_{\mathrm{obj}} / 10^{12} M_\sun > 3 h_{\mathrm{70}}^2 ( r_0 / \unit{1 Mpc})^3
\end{equation}
where $h_{\mathrm{70}}=H_0/70 \unit{km/s/Mpc}$. The paper also gives an estimate of the isolation time,
about 120 Gyr for typical clusters \cite{FutureStructure}.

This expansion dynamics is a key constraint on the ambitions of far-future civilizations. Most matter within not just the observable but the colonizable (at least given assumptions as in \cite{Spamming}) universe will be lost. While no doubt advanced civilizations might wish to affect the expansion rate it seems unlikely that such universal parameters are changeable\footnote{A civilization will by necessity be spatially local, so any change in $\Lambda$ or dark energy parameters would have to be local; beside the inherent problem of how it could be affected, any change would also likely merely propagate at light-speed and hence will not reach indefinitely far. Worse, even a minor change or gradient in the 68.3\% of the total mass-energy represented by dark energy would correspond to massive amounts of normal energy: the destructive effects may be as dramatic as vacuum decay scenarios.}. Depending on the utility function of the civilization, this either implies that there is little extra utility after colonizing the largest achievable bound cluster (utilities dependent on causal connectedness), or that future parts of the civilization will be disconnected from each other (utilities not valuing total connectedness). Civilizations desiring guaranteed separation -- for example, to prevent competitors from invading -- would also value the exponentially growing moats.

Is it possible to gather more mass into gravitationally bound systems? 
In order to move mass, whether a rocket or a galaxy, energy or reaction mass need to be ejected at high velocity to induce motion\footnote{In principle gravitational wave propulsion or spacetime swimming \cite{Wisdom03} are possible alternatives but appear unlikely to be useful in this case.}. This is governed by the relativistic rocket equation $\Delta v = c \tanh( (I_{\mathrm{sp}}/c) \ln(m_0/m_1))$ (we ignore the need for slowing down at arrival). The best possible specific impulse is $I_{\mathrm{sp}}=c$. The remaining mass arriving at the destination will then be $m_1 = m_0 \exp(-\tanh^{-1}(\Delta v/c))$.

In order to overcome the Hubble flow $\Delta v > H_0 r$ (we here ignore that the acceleration of the expansion will require higher velocities). Putting it together, we will get the bound  
\begin{equation}
m(r)  < 4 \pi \rho r^2 \exp(-\tanh^{-1}(H_0 r/c))
\end{equation}
for a concentric shell of radius $r$.

Setting $k=H_0/c$, this can be integrated from $r_0$ (the border of the supercluster) to $1/k$ (the point where it is just barely possible to send back matter at lightspeed):
\begin{equation}
M_{\mathrm{collect}}=(2 \pi \rho/3k^3) \left [ \sqrt{1-k^2x^2}(2k^2x^2-3kx+4)+3\sin^{-1}(kx) \right ]_{\mathrm{r_0}}^{1/k}
\end{equation}

If we use $r_0=50$ Mpc (typical supercluster size) we get $M_{\mathrm{collect}}=3.8\cdot 10^{78} \rho \unit{kg}$, where $\rho$ is the collectable mass density. For $\rho=2.3\cdot 10^{-27} \unit{kg/m^3}$ this is $8.9\cdot 10^{51} \unit{kg}$.
The ratio to mass inside the supercluster (here densities are assumed to be 20 times larger inside) is $M_{\mathrm{collect}}/M_{\mathrm{cluster}} = 12,427$. The collected mass is 35\% of the entire mass inside the reachable volume; the rest is used up.

Another approach would be to convert mass into radiant energy locally, beaming half of it (due to momentum conservation) inwards to a receiver such as a black hole from which it could be extracted later. The main losses would be due to redshift during transmission.

However, this ignores the problem that in order to reach remote locations to send back matter home the civilization needs to travel there. If colonization occurs at lightspeed the colonies will have to deal with an expansion factor $e^{H_0r/c}$ larger, producing the far tighter bound 
\begin{equation}
m(r)<4 \pi \rho r^2 \exp(-\tanh^{-1}(H_0 e^{H_0r/c} r/c)).
\end{equation}
Integrating numerically from $r_0$ to the outer limit $W(1)/k \approx 2.52 \unit{Gpc}$ (where $W$ is Lambert's W function) produces a more modest $M_{\mathrm{collect}}=8.32\cdot 10^{77} \rho$ and $M_{\mathrm{collect}}/M_{\mathrm{cluster}} = 2,705$. Of the total reachable mass, only 7.7\% remains. 

This calculation assumes all mass can be used; if diffuse gas and dark matter cannot be used to power the move, not only does the total mass yield go down by two orders of magnitude but there are going to be significant energy losses in climbing out of cluster potential wells. Nevertheless, it still looks possible to increase the long-term available mass of superclusters by a few orders of magnitude. 

If it is being done it would be very visible, since at least the acceleration phases would convert a fraction of entire galaxies mass-energy into radiation. A radial process would also send this radiation in all outward directions, and backscatter would likely be very noticeable even from the interior.

\subsubsection{Galactic evaporation}
Over long periods stars in galaxies scatter from each other when they have encounters, causing a large fraction to be ejected and the rest are swallowed by the central supermassive black hole. This occurs on a timescale of $10^{19}$ years \cite{Dying}.

However, due to the thermodynamic considerations above, it becomes rational to exploit the universe long before galactic evaporation becomes a problem. 

The same applies to the other forms of long-term deterioration of the universe, such as proton decay, black hole decay, quantum liquefaction of matter etc. \cite{Dying}

\section{Interactions with other civilizations}


The aestivation hypothesis at first appears to suffer the same cultural convergence assumption as many other Fermi question answers: they assume {\em all} sufficiently advanced civilizations -- and members of these civilizations -- will behave in the same way. While some convergence on instrumental goals is likely, convergence strong enough to ensure an answer to the Fermi question
appears implausible since it only takes one unusual civilization (or group within it) {\em anywhere} to break the explanation. Even if it is rational for every intelligent being to do something, this does not guarantee that all intelligent beings are rational.

However, cultural convergence can be enforced. Civilizations could coordinate as a whole to prevent certain behaviors of their own constituents, or of younger civilizations within their sphere of influence. While current humanity shows the tremendous problems inherent in achieving global coordination, coordination may both be important for surviving the emergence of powerful technologies (acting as a filter, leaving mainly coordinated civilizations on the technologically mature side). Even if coordination leading to enforcement of some Fermi question explaining behavior is not guaranteed, we could happen to live inside a domain of a large and old civilization that happens to enforce it (even if there are defectors elsewhere). If such civilizations are very large (as suggested by our intergalactic colonization argument \cite{Spamming}) this would look to us like global convergence.

The aestivation hypothesis strengthens this argument by implying a need for protecting resources during the aestivation period. If a civilization merely aestivates it may find its grip on its domain supplanted by latecomers. Leaving autonomous systems to monitor the domain and preventing activities that decrease its value would be the rational choice. They would ensure that parts of the originating civilization do not start early but also that invaders or young civilizations are stopped from value-decreasing activities. One plausible example would be banning the launch of self-replicating probes to do large-scale colonization. In this scenario cultural convergence is enforced along some dimensions.

It might be objected that devices left behind cannot survive the eons required for restarting the main civilization. While the depths of space are a stable environment that might be benign for devices constructed to be functional there, there are always some micrometeors, cosmic rays or other mishaps. However, having redundant copies greatly reduce the chance of all being destroyed simultaneously. It is possible to show that by slowly adding backup capacity (at a logarithmic rate) a system can ensure a finite probability of enduring for infinite time\footnote{In practice physics places a number of limitations on the durability such as proton decay or quantum tunneling but these limitations are largely outside of the timescales considered in this paper.} \cite{Backups}. The infrastructure left behind could hence be both extremely long-lasting and require a minuscule footprint, even if it is imperfect.


One can make the argument that defenders are likely to win since the amount of materiel they have at home can easily dwarf the amount of materiel that can be moved into place, since long-range transport is expensive. 
However, an interior point in an aestivator domain can be targeted with resources rising quadratically with time as the message goes out. Which effects wins out depends on the relative scaling of the costs and resources\footnote{For example, using the Lanchester square law model of warfare \cite{Lanchester} for a spherical defender domain (of value $\propto r^3$) and quadratically arriving attackers it will resist for time $t \propto \alpha^{1/2}r^{2/3}\eta^{-1/2}$, where $\alpha$ is the defender firepower and $\eta$ is the resource efficiency of transporting attack materiel. The cost to the attacker will scale as $t^3\propto \alpha^{3/2}r^2\eta^{-3/2}$. For sufficiently large $\alpha$ or low $\eta$ it may hence be rational to overlook small emergent civilizations -- or ensure that they do not appear in the first place.}.


One interesting observation is that if we are inside an aestivating civilization, then other aestivators are also likely: the probability of intelligence arising per spacetime hypervolume is high enough that the large civilization will likely encounter external mature civilizations and hence needs to have a strategy against them.

Two mature large-scale civilizations encountering each other will be essentially spherical, expanding at the same rate (set by convergence towards the limits set by physics). At the point of contact the amount of resources available will be the intersection of an expanding communications sphere and the overall colonization sphere: for large civilizations this means that they will have nearly identical resources. Given the maturity assumption they would hence be evenly matched\footnote{The smaller civilization would have a slight disadvantage due to the greater curvature of its surface but if interaction is settled early this might not come into play.}. They have several choices: battle each other for resources, maintain a tangential hyperboloid boundary, or, if their utilities are compatible, join. Given the intention of using resources later, expending them in the present is only rational if it prevents larger losses in the long run. If both civilizations are only interested in resources per se, it may be possible to make any invasion irrational by scorched earth tactics: the enemy will not gain anything from invading and merely expend its resources. This is no guarantee that all possible civilizations will be peaceful vis-\'a-vis each other, since there might be other value considerations (e.g. a negative utilitarian civilization encountering one planning to maintain a large amount of suffering: the first civilization would have a positive utility in reducing the resources of the second as much as possible, even at the expense of future computation). However, given the accelerating expansion maintaining borders could in principle be left to physics.

\section{Discussion}

This paper has shown that very big civilizations can have small footprint by relocating most of their activity to the future.

Of the 6 assumptions of the aestivation hypothesis, 1 (early civilizations) and 2 (broad expansion) are already likely (and if not true, provide alternative answers to the Fermi question). The fifth assumption, that many civilizations wish to aestivate, is supported by the vast increases in computational ability. 

The third assumption, that coordination problems can be resolved, remains hard to judge. 
It should be noted that planning for aestivation is something that is only rational for a civilization once it has solved urgent survival issues. A species that has not reduced its self-generated existential risk enough has good reason to discount the far future due to uncertainty. Conversely, the enormous potential value of a post-aestivation future makes the astronomical waste argument \cite{Waste} stronger: given the higher stakes, the importance of early risk mitigation -- and hence coordination -- increases. 

The fourth assumption is at present also hard to judge. It seems likely that the technology allowing long-range expansion (automation, interplanetary manufacturing, self-replication and long-lived autonomous devices) would enable maintaining arbitrarily large local stockpiles of equipment to fend off incursions. 


Assumption six, the invisibility of aestivation may at first appear hard to test -- nearly any number of advanced civilizations with nearly no energy usage could easily hide somewhere in the galactic halo \cite{Bradbury06}. 
However, in order for it to make sense to aestivate the amount of resources lost during the wait must be small enough that they are dwarfed by the resource costs of efforts to prevent them.

Are there reasons to perform visible megascale engineering to preserve resources? This depends primarily on whether baryonic matter ``construction material'' or mass-energy is regarded as the limiting factor for future computation. If baryonic matter is the key factor considerations of stellar activity, galactic wind or galactic collision mass loss do not seem to imply much utility in megascale engineering to preserve matter. However, if dark matter halos represent significant value (the mass-energy case) reduction of collision loss would be rational, likely doable and very visible. The lack of such galactic engineering hence puts a limit on the existence of such energy-concerned aestivating civilizations in the surveyed universe.

Engineering large-scale galactic movement to prevent separation of superclusters would also be highly visible and the lack of such activity implies that there are no civilizations seeking to optimize long-term causally connected mass-energy concentrations. 

Together, these considerations suggest that if there are aestivating civilizations in our vicinity they are either local (no interest outside their own galaxies or supercluster, utility functions that place no or little value extra matter or energy) or they have utility functions that may drive universal expansion but hold little interest in causal connectedness. 


From a Dysonian SETI perspective, the aestivation hypothesis makes a very clear prediction: look for inhibition of processes that permanently lose matter to inter-cluster or inter-galactic space or look for the gravitationally bound structures more massive than what the standard $\Lambda$CDM cosmology predicts for a given lengthscale.

\subsection{Cosmology/physics assumptions}
What if we are wrong about the model of the universe used in this paper? A few ``nearby'' models have clear implications. In Big Rip scenarios where the scale factor becomes infinite at some point due to phantom energy it is rational to use up available energy before this point. For plausible values of $w$ this is likely far in the future and hence aestivation still makes sense. If there is no horizon radiation, then it is rational to delay until proton or black hole decay, or when the heating due to the civilization becomes on par with the universe. Again aestivation makes sense. 

More fundamentally there is the uncertainty inherent in analysing extreme future scenarios or future technology: even when we base the arguments on well-understood and well-tested physics, there might exist unexpected ways of circumventing this. Unfortunately there is little that can be done about this.

The aestivation hypothesis assumes that the main driver of advanced civilizations is computations whose cost are temperature dependent. More philosophically, what if there are other forms of value that can be generated? Turning energy straight into value without computation would break the temperature dependency, and hence the scenario. 

This suggests an interesting line of investigation: what is the physics of value? Until recently the idea that information was physical (or indeed, a measurable thing) was exotic but currently we are seeing a renaissance of investigations into the connections between computation and physics. The idea that there are bounds set by physics on how much information can be stored and processed by one kilogram of matter is no longer strange. Could there exist similar bounds on how much value one kilogram of matter could embody?

\subsection{Anthropics}
Does the aestivation hypothesis have any anthropic implications? The main consequence of the physical eschatology considerations in this paper is that future computation could vastly outweigh current computation and we should hence expect most observers to exist in the far future rather than the early stelliferous era. 


The Self-Indication Assumption (SIA) states that we should reason as if we were randomly selected from the set of all possible observers \cite{bostrom2013anthropic}. This is normally assumed to support that we are in a world with many observers. We should expect aliens should exist (since a universe with humans and aliens have more observers, especially if the aliens become a very large post-aestivation civilization).
The aestivation hypothesis suggests that initially sparse worlds may have far more observers than worlds that have much activity going on early (and then run out of resources), so the SIA suggests we should believe in the hypothesis. 

The competing Self-Sampling Assumption states that we should reason as if we were randomly selected from the actually existent observers (past, present, future). This gives a more pessimistic outcome, where the doomsday argument suggests that we may not be going to survive. However, both the SIA and SSA may support the view that we are more likely to be a history simulation \cite{Simulation} running in the post-aestivation era (if that era is possible) than the sole early ancestor population.


\subsection{Final words}
The aestivation hypothesis came about as a result of physical eschatology considerations of what the best possible outcome for a civilization would be, not directly an urge to solve the Fermi question. However, it does seem to provide one new possible answer to the question:
\begin{quote}
That is not dead which can eternal lie.\\
And with strange aeons even death may die.\\
H.P. Lovecraft
\end{quote}

\subsection*{Acknowledgments}
We wish to acknowledge Nick Beckstead,Daniel Dewey, Eric Drexler, Carl Frey, Vincent M\"uller, Toby Ord, Andrew Snyder-Beattie, Cecilia Tilli, Owen Cotton-Barratt, Robin Hanson and Carl Shulman for helpful and stimulating discussion. 

\bibliographystyle{ieeetr}
\bibliography{gog}{}

\end{document}